\documentclass[traditabstract,preprint]{aa}
\usepackage{epsfig,graphicx,natbib}
\usepackage{longtable}
\usepackage{amssymb}
\usepackage{amsmath}

\def\PKS1830{\hbox{PKS\,1830$-$211}}

\begin{document}

\title{Dual differential polarimetry. A technique to recover polarimetric information from dual-polarization observations.}

\author{I. Mart\'i-Vidal\inst{1} \and 
        W. H. T. Vlemmings\inst{1} \and
        S. Muller\inst{1}
}

\offprints{I. Mart\'i-Vidal \\ \email{mivan@chalmers.se}}
\institute{Department of Earth and Space Sciences, Chalmers University of Technology, 
           Onsala Space Observatory, SE-43992 Onsala, Sweden
}

\date{Accepted in A\&A.}
\titlerunning{Dual differential polarimetry in interferometric observations}
\authorrunning{Mart\'i-Vidal et al. (2015)}

\abstract{

Current mm/submm interferometers, like the Atacama Large mm/submm Array (ALMA), use receivers that register the sky signal in a linear polarization basis.   
In the case of observations performed in full-polarization mode (where the cross-correlations are computed among all the polarization channels) it is possible to reconstruct the full-polarization brightness distribution of the observed sources, as long as a proper calibration of delay offsets and leakage among polarization channels can be performed. Observations of calibrators, preferably with some linear polarization, with a good parallactic angle coverage are usually needed for such a calibration. In principle, dual-polarization observations only allow us to recover the Stokes $I$ intensity distribution of the sources, regardless of the parallactic angle coverage of the observations.
In this paper, we present a novel technique of dual differential polarimetry that makes it possible to obtain information related to the full-polarization brightness distribution of the observed sources from dual-polarization observations. This technique is inspired in the Earth-rotation polarization synthesis and can be applied even to sources with complex structures.  
}

\keywords{instrumentation: interferometers -- techniques: interferometric -- techniques: polarimetric}

\maketitle

\section{Introduction}

Polarization is an essential source of information in the study of astronomical objects. Polarization can be recorded in two different ways, using receivers with either circular- or linear-polarization feeds. 
At mm/submm wavelengths, new-generation interferometers like ALMA are being equipped with linear polarization feeds, which decouple the sky signal into two streams of linear polarization, one horizontal (with respect to either the antenna axis or the frame of the receiver feed), called $X$, and one vertical, called $Y$. These receivers have a high polarization purity, which does not degrade for wide bandwidths. This is not the case for circular-polarization receivers.

There are, however, some disadvantages of linear-polarization receivers used in interferometric observations, such as a more complex parallactic-angle correction when the baselines involved are very long, or some degeneracies in the calibration when the calibrators are linearly polarized and the parallactic-angle coverage is not good enough (this limitation is also present in circular-polarization feeds, although to a lower extent).

A common calibration and analysis of interferometric observations on a linear-polarization basis only provides full-polarization information of the source structure when the observations are performed in the so-called full-polarization mode. In this mode, cross-correlations of all the polarization channels of the antennas are performed: $XX$, $XY$, $YX$, and $YY$. In principle, dual-polarization observations (where only the cross-products $XX$ and $YY$ are performed) would not be sufficient to extract polarization information of the observed sources as long as we use an ordinary data reduction approach.

Nevertheless, 
it is still possible to obtain some polarimetry information from dual-polarization data. This information is subtly encoded in the relative values of the visibilities in $XX$ and $YY$, and in the time dependence of the parallactic angle, $\psi$, of the antennas. The Earth rotation changes $\psi$ and allows us to synthesize the source polarization distribution, through the changing projection of the source polarization angle, $\phi$ on the receiver axes, $X$ and $Y$. 

The use of the Earth rotation to derive polarimetry information of unresolved (i.e., point-like) sources has been used in single-dish observations (e.g., \citealt{Trippe}). In this paper, we present the new dual differential polarization technique, which makes use of the $\psi$ dependence of the $XX$ and $YY$ interferometric observations, to derive polarimetry information of sources with a resolved structure. The performance of the technique on simulations and real observations is also discussed.

\section{Dual differential polarimetry technique}
\label{TECHNIQUE}

The Stokes parameters $I$, $Q$, $U$, and $V$ encode all the information about the polarization state of light. $I$ is the total intensity (polarized plus unpolarized), $Q$ and $U$ encode the information about the linear polarization, and $V$ is related to the circular polarization. Full-polarization interferometric observations allow us to reconstruct the sky brightness distribution of each of the four Stokes parameters, from which we can derive all the polarimetric information about the observed sources. When linear-polarization feeds are used, the cross-correlations $XX$, $XY$, $YX$, and $YY$ are related to non-redundant linear combinations of $I$, $Q$, $U$, and $V$, where $XX$ and $YY$ are only affected by $I$, $Q$, and $U$. In particular, the $XX$ visibilities are related to the Fourier transform of the brightness distribution of the sum of Stokes parameters $I + Q$ \citep[see, e.g. ][]{smirnov2011a}. In a similar way, the $YY$ visibilities are related to the brightness distribution of $I - Q$. The Stokes parameter $Q$ in these expressions is given in the frame of the receiver (i.e., horizontal and vertical with respect to the receiver frame), which is not to be confused with  the equatorial sky frame (i.e., referring to the celestial meridian that crosses over the source). As a consequence, we must apply a parallactic angle correction to the visibility matrix \citep{smirnov2011a} to obtain the correct relationship between visibilities and brightness distribution of the Stokes parameters. If the baselines of our interferometer are not very long (so that we can assume the same parallactic angle, $\psi$, for all antennas), we can write

\begin{equation}
\begin{pmatrix}  V_{xx} & V_{xy} \\ V_{yx} & V_{yy} \end{pmatrix} = P \begin{pmatrix}  V^c_{xx} & V^c_{xy} \\ V^c_{yx} & V^c_{yy}  \end{pmatrix} P^H,
\label{VisRotEq}
\end{equation}

\noindent where $P$ is the parallactic-angle matrix \citep[e.g.][]{smirnov2011a}, 

\begin{equation}
P = \begin{pmatrix} \cos{\psi} & \sin{\psi} \\ -\sin{\psi} & \cos{\psi} \end{pmatrix},
\label{PMatEq}
\end{equation}

\noindent $P^H$ is the Hermitian of $P$, $V_{kl}$ are the visibility products between polarizations $k$ and $l$, and $V^c_{kl}$ are the visibilities properly corrected for parallactic angle. For the brightness matrix, we have

\begin{equation}
\begin{pmatrix} I+Q & U+iV \\ U-iV & I-Q \end{pmatrix} = P \begin{pmatrix} I+Q' & U'+iV \\ U'-iV & I-Q' \end{pmatrix} P^H,
\label{BrightRotEq}
\end{equation}

\noindent where the Stokes parameters $Q'$ and $U'$ are measured in the frame of the sky (i.e., corrected for parallactic angle), whereas $Q$ and $U$ are given in the frame of the antenna receivers. When there are no direction-dependent calibration effects in the data (this is especially true at mm/submm wavelengths), we also have \citep{smirnov2011a}

\begin{equation}
\begin{pmatrix} V^c_{xx} & V^c_{xy} \\ V^c_{yx} & V^c_{yy} \end{pmatrix} = \int_{\vec{\sigma}}{\begin{pmatrix} I+Q' & U'+iV \\ U'-iV & I-Q' \end{pmatrix}\exp{(-2\pi\frac{\vec{B_{\perp}}\vec{\sigma}}{\lambda})}d\vec{\sigma}},
\label{RIME}
\end{equation}

\noindent where $\vec{\sigma}$ is a unit vector that runs over the source brightness distribution (i.e., $I$, $Q$, $U$, and $V$ are functions of $\vec{\sigma}$) and $\vec{B_{\perp}}$ is the projection of the baseline vector into the {\em uv}-plane (i.e., the plane perpendicular to the source direction). We assume that the visibilities are already calibrated for bandpass and amplitude/phase gains (the effect of polarization leakage is studied in Sect. \ref{LEAKAGE}).

From the equations above, it is straightforward to show that

\begin{equation}
V_{xx} = \int_{\vec{\sigma}}{(I+Q^\psi)\exp{(-2\pi\frac{\vec{B_{\perp}}\vec{\sigma}}{\lambda})}d\vec{\sigma}}
\label{NewQEq0}
\end{equation}

\noindent and

\begin{equation}
V_{yy} = \int_{\vec{\sigma}}{(I-Q^\psi)\exp{(-2\pi\frac{\vec{B_{\perp}}\vec{\sigma}}{\lambda})}d\vec{\sigma}},
\label{NewQEq1}
\end{equation}

\noindent where $Q^\psi = Q'\cos{(2\psi)} - U'\sin{(2\psi)} = Q$. According to these equations, and for small fields of view, the Fourier transform of $V_{xx}$ for a constant parallactic angle, $\psi$, is related to the brightness distribution of $I+Q$, where $Q$ is defined in the frame of the antenna receiver. In a similar way, the Fourier transform of $V_{yy}$ is related to the brightness distribution of $I-Q$. 

\subsection{Case 1. Perfect amplitude calibration}

We now assume that we have a set of interferometric observations in dual-polarization mode for which the parallactic angles of all antennas, $\psi$, are equal and constant. We can reconstruct the brightness distribution of $I+Q^\psi$ and $I-Q^\psi$ by solving the inverse Fourier equation for $V_{xx}$ and $V_{yy}$, respectively. We call $I_{xx}^\psi$ and $I_{yy}^\psi$ the sky brightness distributions corresponding to $V_{xx}$ and $V_{yy}$. 
When $\phi$ is the position angle of the polarization vector (in the frame of the sky) and $p$ is the fractional polarization (both quantities are assumed to vary throughout the extent of the source), we have

\begin{equation}
\frac{I_{xx}^\psi}{I_{yy}^\psi} = \frac{1 + p\cos{\left(2(\phi-\psi)\right)}}{1-p\cos{\left(2(\phi-\psi)\right)}},
\label{RatioEq0}
\end{equation}

\noindent since $Q^\psi = I\,p\,\cos{\left(2(\phi-\psi)\right)}$. In this equation, $I$ is the brightness distribution of the Stokes $I$ parameter. If the visibility amplitudes are perfectly calibrated, the brightness distribution $I$ can be obtained by inverting the Fourier equation for the sum of visibilities $V_{xx}+V_{yy}$. Then, the ratio in Eq. \ref{RatioEq0}, together with $I$, will provide information on the sky distribution of $p$ and $\phi$ throughout the source. In principle, this is an underdetermined problem because we have one constraint (Eq. \ref{RatioEq0}) and two parameters ($p$ and $\phi$) for each source component in $I$. However, if we have observations of the same source under different parallactic angles, we can apply Eq. \ref{RatioEq0} to different values of $\psi$. Hence, we can recover the $p$ and $\phi$ distribution by means of a least-squares minimization applied to each individual component of the source brightness distribution.

\subsection{Case 2. Amplitude bias between $V_{xx}$ and $V_{yy}$}

A more realistic case is an imperfect calibration of the absolute flux scale and/or bandpass for $X$ and $Y$ signals.
As a consequence, frequency-dependent global amplitude offsets may appear between the $XX$ and $YY$ visibilities. This is especially true for dual-polarization observations, since all the amplitude calibration is performed in a completely independent way for the $X$ and $Y$ polarization channels. In these cases, we can still use Eq.~\ref{RatioEq0}, but without computing $I$ from the mere sum of visibilities in $XX$ and $YY$. 

We select a fiducial component in the source brightness distribution (for instance, the phase center, or the peak intensity).  
We call this source component our ``fiducial source element'', FSE. We compute the total flux density of the FSE in the $XX$ and $YY$ images (we call these values $I_{xx0}^\psi$ and $I_{yy0}^\psi$, respectively) and divide the $I_{xx}^\psi$ and $I_{yy}^\psi$ brightness distributions by these values. It is obvious that the ratios $I_{xx}^\psi/I_{xx0}^\psi$ and $I_{yy}^\psi/I_{yy0}^\psi$ are independent of the absolute flux calibration (and also of the amplitude bandpass), since they are related to relative changes in the brightness distribution of the source. These quantities are indeed very robust against calibration artifacts, since their values are mainly encoded in the phase and amplitude closures, which are independent of the antenna gains. The intensity ratios are related to the brightness distribution of Stokes parameters in the following way:

\begin{equation}
\frac{I_{xx}^\psi}{I_{xx0}^\psi} = \frac{I+Q^\psi}{I_0+Q^\psi_0}
\label{RatioEq1}
\end{equation}

and

\begin{equation}
\frac{I_{yy}^\psi}{I_{yy0}^\psi} = \frac{I-Q^\psi}{I_0-Q^\psi_0},
\label{RatioEq2}
\end{equation}

\noindent where $I_0$ and $Q^\psi_0$ are referred to the FSE.
We can now construct a ratio of ratios that depends on $\psi$, which we call polarization ratio, $R_{pol}^\psi$, in the form

\begin{equation}
R_{pol}^\psi = \frac{1}{2}\left(\frac{I_{xx}^\psi/I_{xx0}^\psi}{I_{yy}^\psi/I_{yy0}^\psi}-1\right).
\label{RpolEq0}
\end{equation}

Using Eqs. \ref{RatioEq1} and \ref{RatioEq2}, we obtain

\begin{equation}
2R_{pol}^\psi + 1 =  \frac{I+Q^\psi}{I-Q^\psi}\times\frac{I_0-Q^\psi_0}{I_0+Q^\psi_0}.
\label{RpolEq1}
\end{equation}

If the polarization percentage is small (i.e., $Q<<I$) we can simplify Eq. \ref{RpolEq1} and write

\begin{equation}
R_{pol}^\psi \sim p\cos{\left(2(\phi-\psi)\right)} - p_0\cos{\left(2(\phi_0-\psi)\right)}.
\label{RpolEq1b}
\end{equation}

\noindent In this case, $R_{pol}^\psi$ encodes information related to the difference between the polarization of any component in the source structure and the polarization in the fiducial source element, FSE.

\subsection{Dual differential polarimetry based on Earth rotation synthesis}

Equation \ref{RpolEq1b} (or Eq. \ref{RpolEq1}, if $Q$ is not much smaller than $I$) is the heart of our dual differential polarimetry technique. Given a set of interferometric observations with baselines short enough to assume that $\psi$ is the same for all antennas, we can split the data into snapshots of constant $\psi$. For each snapshot, we reconstruct images of the observed source using the $V_{xx}$ and $V_{yy}$ visibilities. Then, we can compute the sky distribution of $R_{pol}^\psi$ for each parallactic angle, using Eq. \ref{RpolEq0}. From Eq. \ref{RpolEq1b}, the value of $R_{pol}^\psi$ follows the expression

\begin{equation}
R_{pol}^\psi = p_{dif}\,\cos{\left( 2(\phi - \alpha - \psi)  \right)},
\label{RpolEq2}
\end{equation}

\noindent with

\begin{equation}
\alpha = \frac{1}{2}\mathrm{arctan}\left( \frac{p_0\sin{\Delta}}{p - p_0\cos{\Delta}} \right),
\label{alphaEq}
\end{equation}

\begin{equation}
p_{dif} = \sqrt{p^2 + p_0^2 -2p_0p\cos{\Delta}},
\label{pdiffEq}
\end{equation}

\noindent and

\begin{equation}
\Delta = 2(\phi_0 - \phi).
\label{DeltaEq}
\end{equation}

From Eq. \ref{RpolEq2}, we see that $R_{pol}^\psi$ is a sinusoidal function of the parallactic angle $\psi$. The amplitude of the sinusoid is related to the parameter $p_{dif}$ and the argument is related to the parallactic angle and to the source-dependent quantity $\phi - \alpha$. All these parameters depend on the difference of the source polarization between any source component and the FSE. We can derive the sky distribution of differential polarization in a source (i.e., the distribution of $p_{dif}$ and $\phi - \alpha$) by means of a least-squares minimization of $R_{pol}^\psi$ as a function of $\psi$, applied individually to each component of the source structure.

The fitted values of $p_{dif}$ and $\phi - \alpha$ at each source component can then be compared to model predictions of the source, as is usually done in the case of full-polarization observations.

\subsubsection{Implementation of differential polarimetry in Fourier space}
\label{UVANALYSIS}

In the previous section, we have formulated the equations of differential polarimetry based on the relative intensity values of different source components in the image plane. We note, however, that the limited coverage of Fourier space by the interferometer baselines at each snapshot (which determines the shape of the point-spread function, PSF, at each parallactic angle $\psi$), as well as the shorter (snapshot) integration time, may substantially limit the dynamic range (and fidelity) in the images.

There is, however, the possibility of using the complete baseline coverage of the observations (i.e., the PSF corresponding to the full set of observations) in the differential polarimetry analysis, by means of image parametrization techniques. The Stokes I image of the source can be reconstructed using all the available visibilities (i.e., taking advantage of the best possible PSF from the observations). Then, the differential polarization of the components of the (deconvolved) Stokes $I$ source structure can be performed by means of least-squares visibility fitting.

This is the strategy that we have followed in the tests reported in this paper (Sect.\,\ref{TESTS}): we deconvolved the source structures (applying the CLEAN algorithm) using the complete set of observations. Then, we estimated the flux density of each source component, at each parallactic angle, using the visibility-fitting software \texttt{uvmultifit} (\citealt{UVM}). This strategy is similar to performing a compressed sensing analysis at each parallactic angle (see, e.g., \citealt{CS}), but with a null norm-1 normalization Lagrange parameter (since the sparsity in the image model has already been provided by the CLEAN deconvolution of Stokes I).  

This strategy optimizes the fidelity of the source structure being analyzed (given the optimum PSF used in the CLEAN deconvolution) and gives us a reliable estimate of the differential polarimetry signal, because only the flux densities of the source components are fitted at each snapshot (so that the fitted source model is fully linear in visibility space).

\subsection{Effects of Faraday rotation}

When the rotation measure, $RM$, varies across the source, it is possible to measure these variations using $R_{pol}^\psi$, provided that the observed fractional bandwidth is wide enough\footnote{The actual value would depend on the signal-to-noise ratio of the data.}. This strategy was used by \cite{Science} to determine the $RM$ in the jet base of a distant active galactic nucleus, AGN, at submm wavelengths using ALMA observations.

If the rotation measures are different between any pixel of the image and the fiducial source element, Eq. \ref{DeltaEq} takes the form

\begin{equation}
\Delta = 2(\phi_0 - \phi) + 2(RM_0 - RM)\lambda^2,
\label{DeltaEq2}
\end{equation}

\noindent \citep[see appendix of][]{Science} and Eq. \ref{RpolEq2} becomes

\begin{equation}
R_{pol}^\psi = p_{dif}\,\cos{\left( 2(\phi + \lambda^2\,RM - \alpha - \psi)  \right)},
\label{RpolEq3}
\end{equation}

\noindent where $\lambda$ is the wavelength and $RM_0$ is the rotation measure at the fiducial source element. We note that, in this case, the ``synthesis'' of $R^\psi_{pol}$ is not only performed by the different parallactic angles, but also by the different wavelengths of observation. However, it is still required to have a minimum coverage of parallactic angles, regardless of the observed fractional bandwidth, to obtain reliable fits of $R^\psi_{pol}$ using Eq. \ref{RpolEq3}. \cite{Science} observed each epoch under different parallactic angles. This was crucial for them to recover the sinusoidal behavior of $R_{pol}^\psi$ as a function of $\lambda^2$ and $\psi$. 

When $RM_0$ is similar to $RM$, or $\lambda^2\,\left|RM_0 - RM\right| \ll \left|\phi_0 - \phi \right| $, or the fiducial source element is unpolarized (i.e., $p_0=0$), then $\Delta$, $\alpha$, and $p_{dif}$ will be roughly independent of $RM_0$, $RM$, and $\lambda^2$. Thus, Eq.\,\ref{RpolEq3} will only depend on $RM$. This was the assumption used in \cite{Science} to estimate the $RM$ in PKS\,1830$-$211. Given that this source is a gravitational lens, it is plausible to assume that the $RM$ of the lensed images is similar (being one image the fiducial source element), so that $RM \sim RM_0$. 

However, in a more general case, $RM$ and $RM_0$ may differ substantially across the source, so that assuming of a constant $p_{dif}$ and $\alpha$ in Eq.\,\ref{RpolEq3} will bias the estimate of $RM$. Such a bias will depend on the difference between $\phi$ and $\phi_0$ and on the fractional bandwidth of the observations. In Appendix \ref{RMBIAS} we discuss the effects of this bias.

\subsection{Effect of feed rotation at the antennas}

In some cases the X and Y axis of the receivers rotate with respect to the antenna frame. Then, Eq. \ref{RpolEq2} (and all the equations that follow in the next sections) is still valid by adding the feed rotation angle (we call it $\beta$) to the parallactic angle $\psi$. That is, $\psi \rightarrow \psi + \beta$. Then, Eq. \ref{RpolEq2} becomes

\begin{equation}
R_{pol}^\psi = p_{dif}\,\cos{\left( 2(\phi - \alpha - \psi - \beta)  \right)},
\label{RpolEq2b}
\end{equation}

\noindent where the effect of $\beta$ is that of a constant angle added to the cosine. Hence, the dependence of $R_{pol}$ on the intrinsic source polarization (given by both $\alpha$ and $p_{dif}$) is unaffected by $\beta$. In Sect. \ref{PROBLEM} we discuss the effect of the feed rotation in the ALMA antennas on the results recently reported in \cite{Science}.

\subsection{Effect of an off-axis receiver}
\label{SQUINT}

When the receiver horns are not aligned with the axis of the antenna mounts, each polarization channel will have a slightly different beam pattern on the sky. This is related to the well-known beam squint effect and is present in all the ALMA antennas. Hence, for the observations of extended sources, the off-axis receivers of ALMA could introduce a fake differential polarimetry signal accross the source structure. 

We note that the beam squint effect is partially smeared out during the image reconstruction from the visibilities as long as the observations cover different parallactic angles. The reason for this smearing is that the beam squint rotates with the antenna mounts (being thus a function of the parallactic angle). However, the feed rotation with the parallactic angle may imprint a spurious signal in the differential polarimetry analysis. 
We note, however, that given the finite pointing accuracy of the antennas, a fraction of the beam squint effect (over the whole interferometer) will also be averaged out at each integration time, since each antenna will see a slightly different DDE pattern across the source. For the ALMA antennas (and according to the ALMA technical handbook\footnote{https://almascience.eso.org/proposing/call-for-proposals/technical-handbook}), the pointing accuracy is about 0.6''. On the other hand, estimates of the beam squint (although in a circular polarization basis) are given in \cite{ALMA362} (Table VI), and are 
of the order of a few \% of the beam size (2.5\% at most). 

Even if the ALMA antennas had an infinite pointing accuracy,  
the effects of beam squint (using the values in \citealt{ALMA362}) would map into a maximum instrumental $R_{pol}$ up to 1\% at a distance of one-third of the FWHM from the beam center (i.e., from $\sim$20$''$ in band 3 to $\sim$3$''$ in band 9). In the observations of PKS\,1830$-$211 reported in \cite{Science}, this effect (given the small separation among the two lensed images) would fall below 0.15\%, which is several times lower than the signal detected from the source.

We must also note that if there is a time evolution in the polarized source structure, the value of $R_{pol}$ at a given parallactic angle evolves in time. However, if the beam squint effects were to dominate the $R_{pol}$ signal, all the $R_{pol}$ measurements would only depend on the parallactic angle. \cite{Science} reported a time evolution in the differential polarization of PKS\,1830$-$211 based on comparing $R_{pol}$ values at similar parallactic angles and different epochs. 
Such a time evolution would be impossible to explain in terms of effects related to the mere antenna/receiver geometry.

Another effect related to beam squint would be a spurious rotation measure that is due to the frequency-dependent beam pattern. However, this effect is negligible at high frequencies. For the observations reported in \cite{Science} at bands 6 and 7, the spurious frequency dependence in $R_{pol}$ for any given tuning would be below 0.05\%, whereas the frequency dependence of $R_{pol}$ reported in \cite{Science} was, in some epochs, about 100\%.

\subsection{Effects of polarization leakage}
\label{LEAKAGE}

The estimate of $R^\psi_{pol}$ is, by construction, insensitive to bandpass and/or gain differences of the X and Y signals.
The only calibration effect that can change the value of $R_{pol}$ is the polarization leakage in the antenna receivers and any cross-polarization delay or phase. However, the effect of polarization leakage and cross-delays on the $XX$ and $YY$ products is very small (second-order corrections) compared to the effect on the $XY$ and $YX$ products (first-order corrections). 
\cite{smirnov2011a} reported that the Jones matrix for the correction of leakage and cross-phase (or cross-delay) of the $X$ and $Y$ signals in an antenna is

\begin{equation}
J = \begin{bmatrix}1 & 0 \\0 & K\end{bmatrix}\times\begin{bmatrix}1 & D_x \\D_y & 1\end{bmatrix} = \begin{bmatrix}1 & D_x \\D_yK & K\end{bmatrix},
\label{JonesAll}
\end{equation}

\noindent where $K$ is a phase-like factor and $D_x$ and $D_y$ are the complex D-terms that model the polarization leakage in the antenna receivers.  
Since we do not have cross-polarization products in dual-polarization observations, it is not possible to solve for $K$ in the calibration (i.e., to separate it from the phase gains), but, in any case, its effects on the dual-polarization visibilities will not be different from a phase added to the $YY$ product. The observed $XX$ and $YY$ visibilities for the baseline formed by a pair of antennas $A$ and $B$ will be

\begin{equation}
V_{xx}^{obs} = (D^A_x V_{yy}+V_{xy})(D^*)^B_y + D^A_xV_{yx}+V_{xx}
\label{VxxObs}
\end{equation}

and 

\begin{equation}
V_{yy}^{obs} = \left((D^A_y V_{xx} + V_{yx})(D^*)^B_x + D^A_yV_{xy}+V_{yy}\right)K^A(K^*)^B.
\label{VyyObs}
\end{equation}

It is clearly seen that Eq. \ref{VxxObs} and \ref{VyyObs} are basically symmetric to each other, with the exception of a global phase-like factor, $K^A(K^*)^B$ (i.e., the difference of X-Y cross-delays of antennas $A$ and $B$) that will be fully absorbed in the ordinary phase-gain calibration of the $YY$ visibilities.

\begin{figure}[th!]
\centering
\includegraphics[width=9cm]{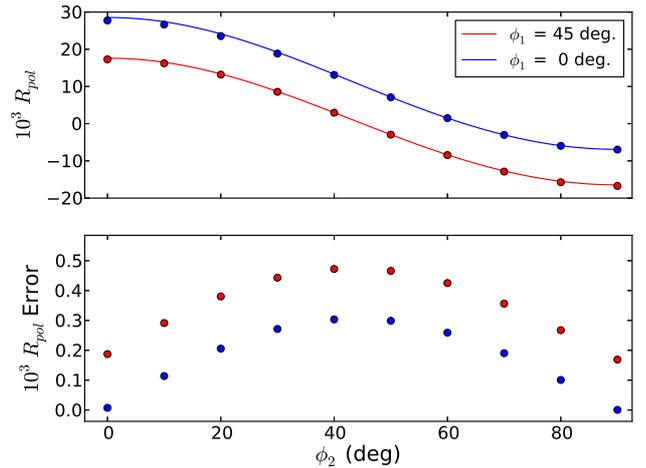}
\caption{Simulated $R^\psi_{pol}$, taking into account effects from polarization leakage in the antenna receivers. Top panel: fitted $R_{pol}$ values (circles) and the model predictions computed from Eq.~\ref{RpolEq1} (solid lines). Bottom panel: difference in $R^\psi_{pol}$ between the fitted values and the model predictions.}
\label{SimulLeak}
\end{figure}

We have simulated ALMA observations of two compact polarized sources located in the same field of view and checked the reliability of $R^\psi_{pol}$ against polarization leakage in the antenna receivers. The two sources were separated by two synthesized beams, their intensity ratio was set to 1.4, and their linear polarizations were 1\% (with a position angle, $\phi_1$, of either 0 or 45 degrees) and 1.7\% (with a position angle, $\phi_2$, from 0 to 90 degrees, in steps of 10 degrees), respectively. No Faraday rotation was introduced in the sources. We simulated ten antennas with leakages of similar amplitudes and random phases. The amplitude leakage used was 1\% (for ALMA, the leakage level in the antenna receivers is about 1\%, according to the ALMA technical handbook). All simulated results are shown in Fig.~\ref{SimulLeak}. All measured values of $R^\psi_{pol}$ follow the prediction of Eq.~\ref{RpolEq1}. It is also clear that $R^\psi_{pol}$ is quite insensitive to the antenna leakage. All errors are below 5$\times$10$^{-4}$. The maximum errors are obtained when the Stokes $U$ signal is strongest in both sources (i.e., the polarization angles of both sources are close to 45 degrees). This is an expected result because the leakage from $V_{xy}$ and $V_{yx}$ into $V_{xx}$ and $V_{yy}$ is larger for a stronger signal in Stokes $U$.

In these simulations, we used a random phase distribution for the D-terms, between 0 and 45 degrees, also adding a correlation between phases of $D_x$ and $D_y$ for each antenna. It is expected that the phase of one should be 180 degrees shifted with respect to that of the other. We note that even lower errors in $R^\psi_{pol}$ would be obtained if the phase distribution of D-terms were wider and/or there were no correlations between $D_x$ and $D_y$ in each antenna.

\section{Testing the techique}
\label{TESTS}

\subsection{Simulations}

We tested the performance of our dual differential polarimetry technique using synthetic data with realistic noise and potential calibration artifacts. In all our simulations, we used the Common Astronomy Software Applications (CASA) package of the National Radio Astronomy Observatory (NRAO)\footnote{\texttt{http://casa.nrao.edu}}. The synthetic data were generated with an in-house modified version of the CASA task \texttt{simobserve}. In this version, full-polarization synthetic datasets can be created for sources with a generic polarization structure. The parallactic angle correction and effects of polarization leakage in the antenna receivers are taken into account. 

Our simulation is based on the ALMA full array configuration ``alma.out04.cfg'', given in the CASA database. The observing frequency was set to 100\,GHz, with a bandwidth of 10\,GHz. An arbitrary source, located at a declination of $-21$ degrees, was observed from an hour angle of $-0.84$\,h to 0.84\,h, resulting in a parallactic angle coverage of between 100 and 260 degrees. The synthesized beam (using natural weighting) was 2.86$\times$2.50\,arcsec, with a position angle, PA, of 89\,degrees. The largest rcoverable scale (LRS) is estimated to be $\sim60''$. In these simulations, we accounted for thermal noise at the receivers, random amplitude errors of up to 5\% in each antenna (and independent for $X$ and $Y$), random phase offsets of up to 10\,degrees between $X$ and $Y$, and random leakage of up to 2\% with random phases of up to 20\,degrees. 

We simulated two cases: 1) an extended unpolarized brightness distribution with an added compact polarized component, and 2) 
an extended polarized structure. We discuss these two simulations in the next subsections. 

\subsubsection{Extended unpolarized plus compact polarized emission}

In this case, the dual differential polarimetry was performed on the compact polarized component, and we used the whole unpolarized brightness distribution as our fiducial source element. 
To simulate the unpolarized extended emission, we used the FITS image of galaxy M\,51 called ``M51ha.fits'', which is commonly used in CASA-related simulations\footnote{\texttt{http://casaguides.nrao.edu/}}. The brightness peak of the model was set to 1\,mJy/pixel, which maps into a peak intensity of 1.35\,Jy/beam in the convolved image. The overall size of the source was set to $\sim 20''\times 30''$.
The compact polarized component was located at about 6$''$ south of the galaxy center, with a flux density of 1\,Jy, a fractional polarization of 8\%, and a polarization angle of $-80$\,degrees. 

\begin{figure}[th!]
\centering
\includegraphics[width=9cm]{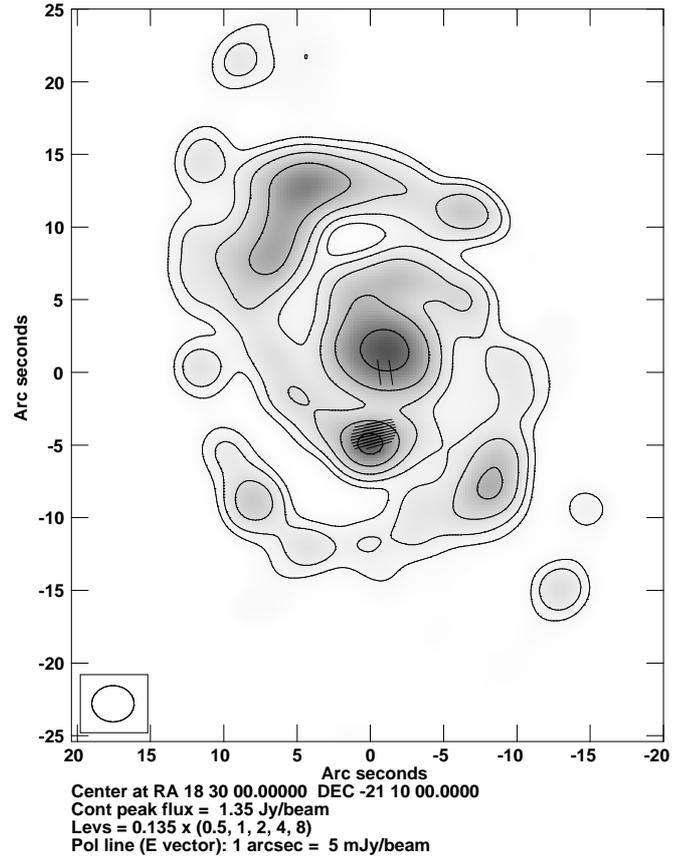}
\caption{CLEANed full-polarization image of our M51 simulation. Vector lines are parallel to the electric field. Note the compact polarized component at the center of the relative coordinates in the image. Owing to leakage effects, there is also a marginal spurious polarization close to the center of the galaxy.}
\label{Simulations}
\end{figure}

A full-polarization image, shown in Fig. \ref{Simulations}, was obtained using an standard CLEAN deconvolution with no self-calibration. The polarized source is clearly detected, together with a weak spurious polarization signal close to the galaxy center, likely due to the polarization leakage.

As we explained in Sect.\,\ref{UVANALYSIS},  
we made use of the \texttt{uvmultifit} program for CASA, a tool for fitting source models to interferometric visibilities \citep{UVM}. We fit a source consisting of two model components to the data. One of the components corresponds to the complete set of CLEAN points found in the deconvolution, but removing those within 1$''$ from the compact polarized source. The second component consists of the set of CLEAN points removed from the first component. 

\begin{figure}[th!]
\centering
\includegraphics[width=10cm]{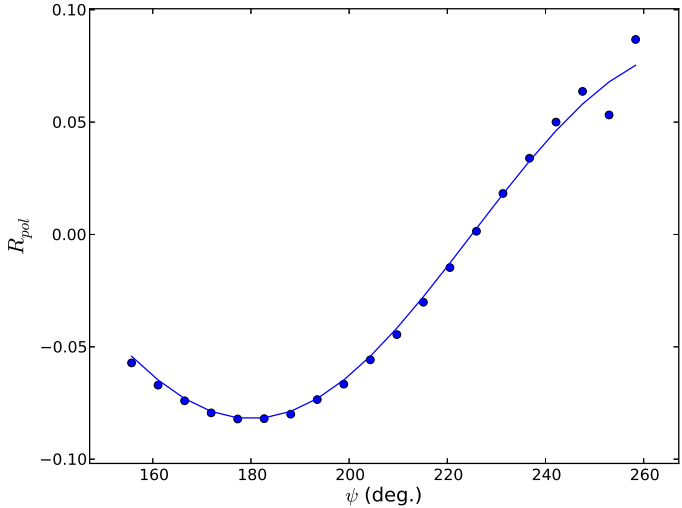}
\caption{$R_{pol}$ between the total unpolarized intensity and the compact polarized source in our modified model of M51. The points are measurements obtained with our software \texttt{uvmultifit}.  The line is the model prediction of $R_{pol}$ for this source.}
\label{Simulations3}
\end{figure}

In our modeling, we left the positions of all the CLEAN points fixed, and only fit the total flux density of each model component (i.e., the overall unpolarized flux for the first component, and the overall polarized flux for the second component) in both $XX$ and $YY$. We separated the data into scans of constant parallactic angle. From the fit flux densities at each parallactic angle, we computed $R_{pol}$ as a function of time throughout the experiment. We show the resulting $R_{pol}$ measurements in Fig. \ref{Simulations3} together with the prediction using Eq. \ref{RpolEq2}. The agreement is excellent. Given that the extended structure is unpolarized, we can recover the full polarization state of the compact source from our measured $R_{pol}$. In this case, $p_0$ in Eq. \ref{RpolEq1b} is exactly zero, so that $R_{pol}$ vs. $\psi$ encodes the information about the absolute (i.e., not differential) polarization, $p$ and $\phi$.

\subsubsection{Extended polarized emission}

An extended polarized source was generated using four elliptical Gaussian intensity distributions. One, with a full width at half maximum (FWHM) of $25'' \times 8''$, accounted for the unpolarized emission. Another with a FWHM $20''\times 6.5''$, accounted for the Stokes $U$ emission, and the other two, with a FWHM of $20''\times 6.5''$, accounted for Stokes $Q$ (we note that the LRS is $\sim60''$). All the position angles were set to 90 degrees (i.e., major axis in the east-west direction). The Gaussians for Stokes $I$ and $U$ were centered on the same position and the Gaussians for Stokes $Q$ were offset by 5\,arcsec, one to the east and the other to the west.

\begin{figure*}[th!]
\centering
\includegraphics[width=18cm]{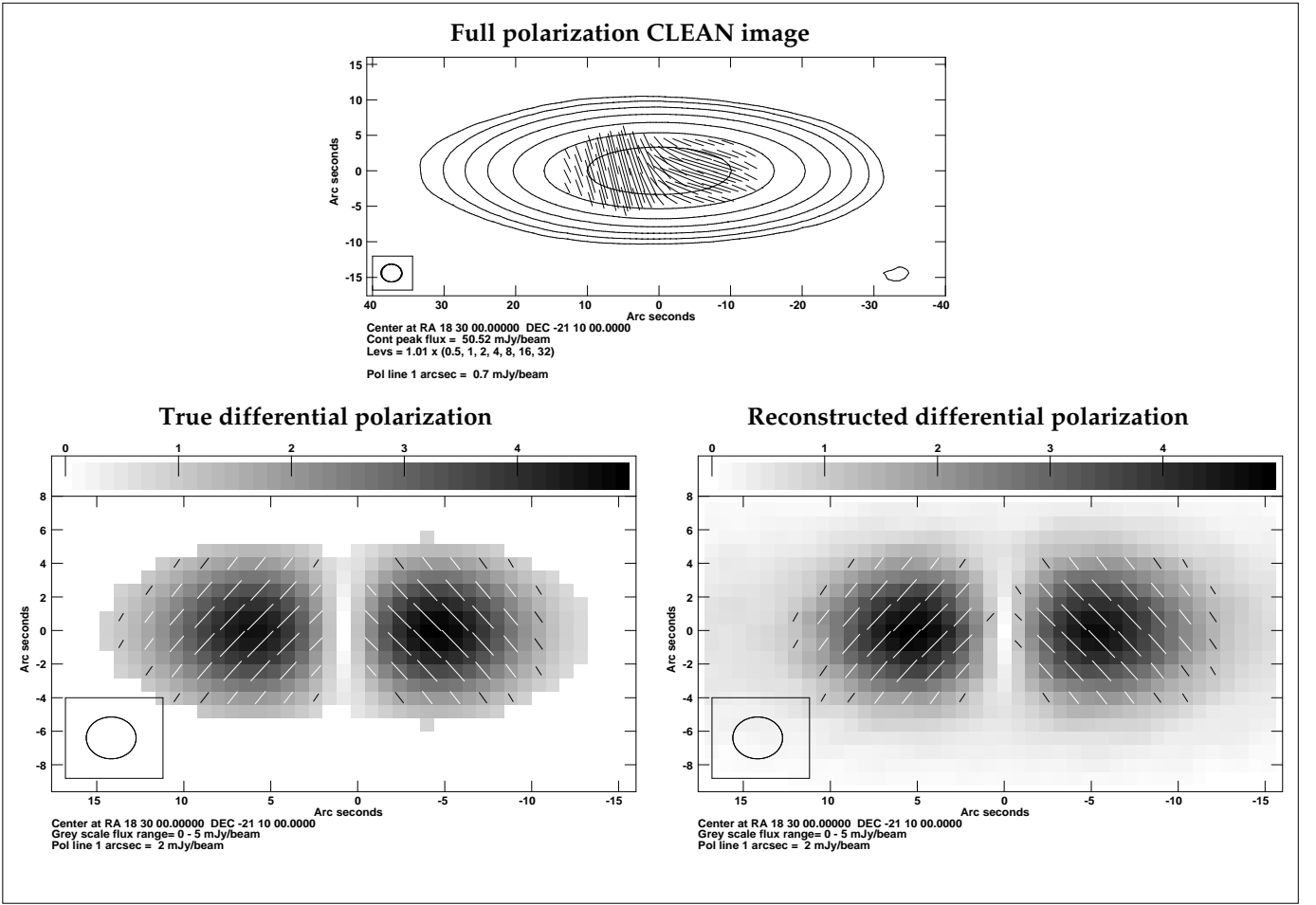}
\caption{Top panel: full-polarization CLEANed image from our simulated dataset of a source with extended polarization. Bottom left panel: distribution of the differential polarization, computed from the source model used in the simulation convolved with the CLEAN restoring beam. Bottom right panel: reconstruction of the differential polarization from the dual-polarization data, using our fitting approach (see text).}
\label{Simulations1}
\end{figure*}

We applied an ordinary deconvolution to the whole full-polarization dataset and obtained the image shown in Fig. \ref{Simulations1} (top). The polarization angle clearly swings by $\sim$90\,degrees in the east-west direction, with a minimum of polarized intensity corresponding to the peak intensity of Stokes $I$. At the peak intensity, the polarization angle is about 45\,degrees.

The analysis of dual differential polarimetry for this source was also performed with the \texttt{uvmultifit} program. We CLEANed Stokes $I$ (with no polarization calibration applied) and took each CLEAN delta component as an independent model component for the fit. The positions of all the components were fixed, and we only fit the flux densities. All the CLEAN components were simultaneously fit at each scan (i.e., for each parallactic angle). The fits were performed separately for the $XX$ and $YY$ visibilities. The initial flux-density values used in the fit were set to those obtained by CLEAN in the deconvolution of Stokes $I$.

\begin{figure}[th!]
\centering
\includegraphics[width=10cm]{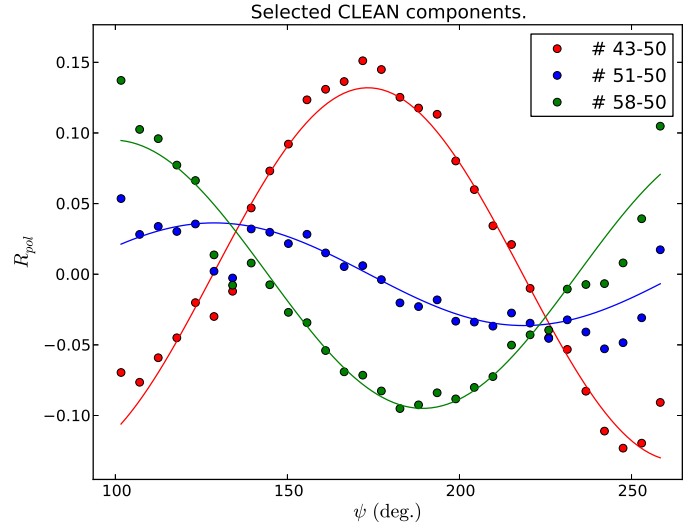}
\caption{$R_{pol}$ as a function of parallactic angle for three selected pixels in the source image with extended polarization. Each color corresponds to a different pixel. The numbers given in the legend are the X and Y pixel coordinates in the image. The $R_{pol}$ values were obtained with our software \texttt{uvmultifit}. Lines are fits using Eq. \ref{RpolEq2}.}
\label{Simulations2}
\end{figure}

The fit provided a set of flux densities for each CLEAN component, polarization product ($XX$ and $YY$), and parallactic angle. Then, for each parallactic angle, we convolved the best-fit models for $XX$ and $YY$ with the CLEAN restoring beam, and scaled the resulting images to their values at the peak intensity of Stokes I (i.e., $I^\psi_{xx0}$ and $I^\psi_{yy0}$ in Eq. \ref{RpolEq0}). Thus, the peak intensity is our FSE. Then, we computed $R^\psi_{pol}$ pixel by pixel for each parallactic angle and fit the results at each pixel to the model given in Eq. \ref{RpolEq2}. We show some example results of these fits in Fig. \ref{Simulations2}. From these fits, we obtained the value of $p_{dif}$ and $\phi - \alpha$ for each pixel. These results are shown in Fig. \ref{Simulations1} (bottom right). Using the model Gaussians for Stokes $I$, $Q$, and $U$, we can also compute the true dual differential polarimetry of the source (Fig. \ref{Simulations1}, bottom left), and compare it to our image reconstruction. The swing in the polarization angle is clearly seen in the image reconstruction and the true model. The fractional polarization (gray scale) in our reconstruction is also very similar to the model, demonstrating the reliability of our differential polarization method.

\subsection{Test with ALMA Science Verification data}
\label{DATA}

The first released Science Verification (SV) full-polarization ALMA dataset consists of observations of source 3C\,286 in band 6 (hereafter, we just refer to the central frequency in the observations, 225\,GHz). The full description of the data and the calibration procedure can be found in the CASA Guides database\footnote{\texttt{https://casaguides.nrao.edu/index.php/3C286\_Polarization}} and in \cite{Nagai}. We used these SV observations to test our technique, by comparing our results to those published in the CASA Guide.

Given that source 3C\,286 is only barely resolved with ALMA, a first impression is that it is probably not possible to test the differential polarization algorithm with this dataset. However, we can still use this technique to derive the absolute source polarization if we use one additional assumption, which we describe below. 

Instead of using a given portion of the target source as the fiducial source element, FSE, we used the absolute flux calibrator (Ceres). Assigning the role of the FSE to a different source requires that either the parallactic angles of both sources (reference and target) are the same throughout the observations (which is unlikely to happen for any pair of sources on the sky) or that the reference source is unpolarized (so that the XX and YY visibilities are always the same, regardless of the parallactic angle). 

The absolute flux calibrator was used to scale the XX and YY visibilities to their correct amplitude values. Thus, if we assume that the overall ratio of gain amplitudes in the array does not change significantly throughout the observations, we can derive the $R_{pol}$ of the target from a simple visibility fitting of the target flux densities in XX and YY (as a function of the parallactic angle). The details of this analysis are given in the next paragraph.

We performed the bandpass and gain calibration as described in the CASA Guide of this SV dataset, with the exception that we applied the same amplitude solution to the two polarizations, thus ensuring that any polarization signal from the phase calibrator is not transferred to the target (gain type ``T'' in the CASA task \texttt{gaincal}). When the absolute flux calibrator (Ceres) was used to scale the amplitude gains to their respective values, our calibration strategy diverged from what is described in the Guide. We applied the phase and amplitude gains of the phase calibrator to the target and split the data for imaging. Obviously, no parallactic angle correction was applied in the calibration.

\begin{figure}[th!]
\centering
\includegraphics[width=9.5cm]{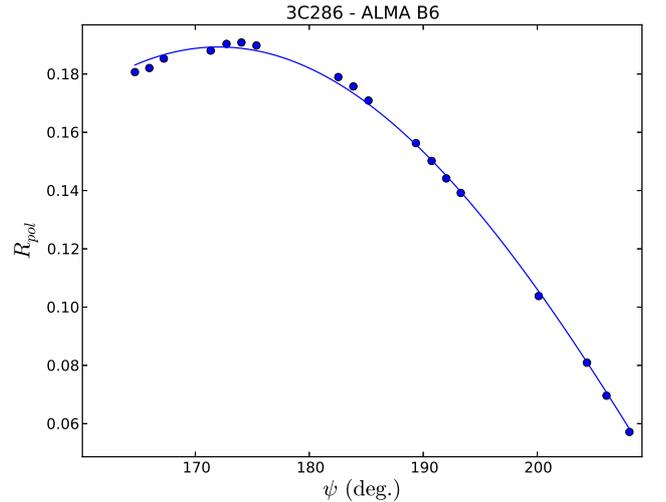}
\caption{$R_{pol}$ of 3C\,286 at 225\,GHz, from ALMA SV observations. The error bars are smaller than the symbol sizes.}
\label{ALMASV}
\end{figure}

We note that at this stage, no polarization-related calibration was applied to the data (neither X-Y cross-phase nor leakage). Two iterations of phase self-calibration and one iteration of amplitude self-calibration (in ``T'' gain type mode) were applied to the target, using Stokes $I$ for the imaging. Finally, we used \texttt{uvmultifit} to estimate the flux density of the source for each scan and polarization product (XX and YY), using a point source (fixed at the field center) as fitting model. The resulting $R_{pol}$ values together with the fit to a cosine function are shown in Fig. \ref{ALMASV}.

Our best-fit fractional polarization is $p = 0.189  \pm  0.001$, which is similar to the one reported in the CASA Guide ($p = 0.1645  \pm  0.0002$). We note that our true uncertainty may be larger than the \texttt{uvmultifit} estimate because we assumed a constant overall gain amplitude ratio throughout the observations, and there might be systematics related to the uncertainty in the absolute flux density calibration over the two polarization products. 

In regard to the polarization angle, we must note that the band 6 feeds at ALMA are rotated $-45$\,degrees with respect to the antenna axes (this information is provided in the metadata of the observations). Adding this feed rotation to our angle budget (see Eq. \ref{RpolEq2b}), we obtain a source polarization angle of $\phi = 37.04 \pm 0.12$\,deg., which is consistent with the value reported in the CASA Guide ($38.30 \pm 0.02$\,deg.).

We have thus shown that our method can be used even on compact sources, as long as an unpolarized source is also observed and the overall amplitude ratio of the antenna gains does not change significantly throughout the observations (or between consecutive observations of an unpolarized source).

\section{Reanalysis of Mart\'i-Vidal et al. (2015)}
\label{PROBLEM}

\cite{Science} reported the first practical application of the dual differential polarimetry technique. Their analysis was performed on the lensed images of the gravitationally lensed blazar PKS\,1830$-$211 (with a separation of 1$''$). Using one of the lensed images as the polarization reference, the authors found clear signatures of a frequency dependence on $R_{pol}$ for each epoch and parallactic angle. For a subset of epochs close by in time (at most, a difference of two days was allowed) where the spectral coverage was wide enough, we successfully fit Eq. \ref{RpolEq3} to the data at bands 6 and 7, finding high values of $RM$ ($\sim10^7$ in the observer frame). These values were indicative of strong magnetic fields in the region close to the jet base.

At the time of the observations reported in \cite{Science}, the corresponding metadata provided by the ALMA observatory did not contain correct information about the feed rotation of each ALMA receiver. In addition to this, the true rotation angle of the band 7 receivers, currently provided by the observatory, contradicts the information that can be extracted from the currently available documentation related to the antenna optics (ALMA Memo \#362). In the former, a value of $-53.5$\,degrees is obtained, while in the latter a value of $0$\,degrees is expected.

\begin{figure*}[th!]
\centering
\includegraphics[width=17cm]{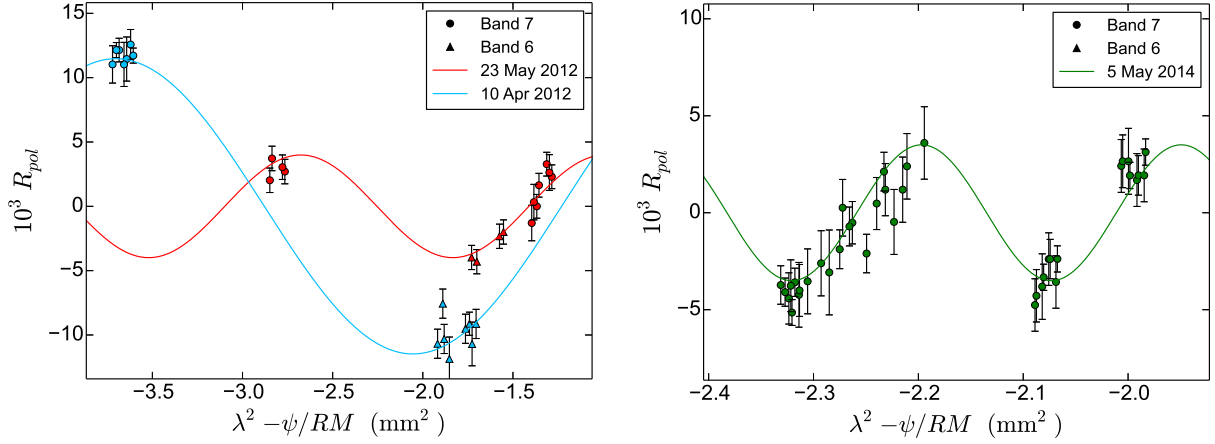}
\caption{$R_{pol}$ between the lensed images of PKS\,1830$-$211. Data taken from \cite{Science}, but applying the feed rotation angles of the ALMA receivers, as currently provided by the observatory in the ALMA metadata.}
\label{NEWSCI}
\end{figure*}

The fact that the feed angles are different between bands 6 and 7 might introduce a bias in the $RM$ reported in \cite{Science}. Fortunately, this bias is rather small and does not affect the main scientific results. In Fig. \ref{NEWSCI} we show the results of a reanalysis of the data reported in \cite{Science}, but using the correct feed angles. We note that the large $RM$s found by \cite{Science} imply a lower effect of the parallactic angle $\psi$ (and hence the feed angle $\beta$), compared to that of the spectral coverage $\lambda^2$. The effect of the $RM$ on the observations of PKS\,1830$-$211 is strong enough to be observable even within the spectral windows of each individual ALMA tuning.  

The new $RM$s determined for epochs 10 April 2012 and 23 May 2012 are $(2.0\pm1.0)\times10^6$\,rad/m$^2$ and  $(3.7\pm0.5)\times10^6$\,rad/m$^2$ (observer's frame), respectively. These are 4.5 and 2.5 times lower than those reported in \cite{Science}, although these new results do not affect the main scientific conclusions of that paper. Regarding the epoch on 5 May 2014, all the observations were performed in band 7, so that the different feed rotation between the two bands does not affect the result reported in \cite{Science}.

\section{Conclusions}
\label{CONCLUSIONS}

We presented the technique of dual differential polarimetry, which is suitable for interferometers whose antennas are equipped with linear-polarization receivers. This technique allowed us to obtain polarimetric information from dual-polarization observations, where only the $XX$ and $YY$ cross-products of antennas are computed. The limitations due to the lack of cross-polarization data, $XY$ and $YX$, can be partially overcome thanks to the Earth rotation synthesis, as long as the baselines of the interferometer are short enough to ensure that all telescopes observe under the same parallactic angle at any time.

Using Earth rotation, we showed that the polarization ratio observable, $R_{pol}$, defined for each pixel in the source image, is a sinusoidal function of the parallactic angle. The amplitude and phase of this sinusoid encodes information about the difference in polarization states between any point in the source and what we call a fiducial source element, FSE (i.e., a reference in the source structure). If the FSE is unpolarized, $R_{pol}$ encodes information about the absolute polarization state of all other regions in the image. If the FSE is polarized, then $R_{pol}$ gives only differential polarization information.  
For observations with wide fractional bandwidths, $R_{pol}$ also encodes information about the rotation measure across the image. In this case, $R_{pol}$ is also a sinusoidal function of the observing wavelength squared, $\lambda^2$.

We tested our technique with realistic simulations of ALMA observations for two case studies: 1) an extended unpolarized source with a compact polarized feature, and 2) an extended polarized source. In both simulations, we were able to reconstruct the sky distribution of differential polarization using the dual-polarization data. 

We also tested our technique using real Science Verification ALMA observations (source 3C\,286 at 225\,GHz) in full-polarization mode. In this case, our technique was also shown to be useful for compact sources, as long as an unpolarized source (or an absolute flux calibrator) has been observed and the overall ratio of gain the amplitudes for all the antennas remains stable throughout the observations.

In the future, this technique will be useful to retrieve polarization information, for instance, from the growing amount of ALMA archival observations that were not designed a priori for polarization purposes.

\begin{acknowledgements}
This paper makes use of the following ALMA data: ADS/JAO.ALMA\#2011.0.00017.SV. ALMA is a partnership of ESO (representing its member states), NSF (USA) and NINS (Japan), together with NRC (Canada), NSC and ASIAA (Taiwan), and KASI (Republic of Korea), in cooperation with the Republic of Chile. The Joint ALMA Observatory is operated by ESO, AUI/NRAO and NAOJ.
\end{acknowledgements}

\begin{appendix}
\section{Bias in the $RM$ differential imaging}
\label{RMBIAS}

\begin{figure*}[th!]
\centering
\includegraphics[width=17cm]{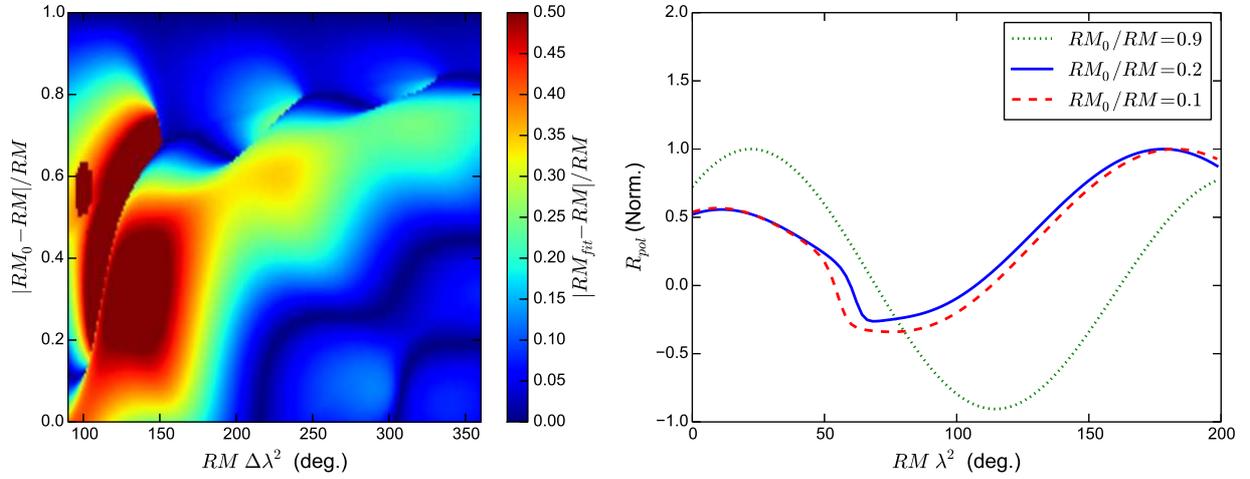}
\caption{Left panel: relative error in the fit rotation measure as a function of the fractional bandwidth (horizontal axis) and the fractional difference (vertical axis) between $RM$ and $RM_0$ (where $RM_0$ is the rotation measure of the fiducial source element). Right panel: some example values of $R_{pol}$ as a function of $\lambda^2$ for different values of $RM_0/RM$. A value of $\phi - \phi_0 = 45^\circ$ has been used.}
\label{RMBias}
\end{figure*}

In Fig.\,\ref{RMBias} we show an estimate of the bias in $RM$, as estimated from the fit of the sinusoid given in Eq.\,\ref{RpolEq3} to the $R_{pol}$ generated accounting for the varying $\alpha$, $\Delta$, and $p_{dif}$ as functions of $RM$, $RM_0$, and $\lambda$. 
We computed $R_{pol}$ for 100 different values of $\lambda$. The range of $\lambda$ was set from $0$ to $\Delta \lambda^2 = \theta_{max}/RM$, where $\theta_{max}$ is the fractional squared-wavelength coverage, normalized to the $RM$ (i.e., $\theta_{max}$ is an angle). In each fit, we assumed a constant $p_{dif}$ and $\alpha$ to introduce the bias in the $RM$ estimate.

To avoid the singularity for $RM=RM_0$ (only when $\phi=\phi_0$), we set $\phi - \phi_0 = 45^\circ$. We also simulated possible depolarization effects by setting $p = p_0\left(\frac{RM_0}{RM}\right)^\gamma$. We let $\gamma$ vary from 0.1 (i.e., almost no depolarization) to 1 (i.e., strong depolarization). The resulting biases in $RM$ are qualitatively similar for the different values of $\gamma$. 

Figure\,\ref{RMBias}\,(left) shows that the bias in $RM$ is smaller when $RM$ and $RM_0$ are very similar (i.e., upper part in the figure). This is an expected result.  
The largest biases in $RM$ are related to fits using relatively narrow fractional bandwidths (i.e., left part in the figure). In these cases, the biases also depend quite strongly on $\phi - \phi_0$ (i.e., the intrinsic angle difference). However, for relatively large fractional bandwidths, where the frequency dependence of $R_{pol}$ covers a good fraction of a $2\pi$ cycle, the biases are smaller (right part of figure).

The values of $R_{pol}$ as a function of $\lambda^2$ are shown in Fig.\,\ref{RMBias}\,(right) for three different cases of $RM_0/RM$. For $RM_0/RM\sim1$, a sinusoid is roughly recovered, whereas strong departures of the sinusoidal shape are seen for larger differences between $RM_0$ and $RM$.

In any case, the bias can be as large as $\sim30$\%, even with wide fractional bandwidths. It is thus desirable that the differential $RM$ imaging is performed either using an unpolarized fiducial source element (i.e., $p_0=0$, so there is no bias) or by fitting Eq.\,\ref{RpolEq3} exactly (i.e., accounting for the $\lambda$ dependence in $p_{dif}$ and $\alpha$). In the latter case, the parameter $RM_0$ would be common in all the fits to the source components, so that a combined fit of the whole source structure would be needed.

\end{appendix}


\begin{thebibliography}{99}


\bibitem[Lamb et al.(2001)]{ALMA362} Lamb, J. W., Baryshev, A., Carter, M. C., et al. 2001, {\em ALMA Memo 362} 

\bibitem[Li et al.(2011)]{CS} Li, F., Cornwell, T.~J., \& de Hoog, F.\ 2011, \aap, 528, A31

\bibitem[Mart\'i-Vidal et al.(2014)]{UVM} Mart\'i-Vidal, I., Vlemmings, W.~H.~T., Muller, S., \& Casey, S. 2014, A\&A, 563, A136 


\bibitem[Mart\'i-Vidal et al.(2015)]{Science} Mart\'i-Vidal, I., Muller, S., Vlemmings, W.~H.~Y., et al. 2015, Science, 348, 311

\bibitem[Nagai et al.(2016)]{Nagai} Nagai, H., Nakanishi, K., Paladino, R., et al.\ 2016, ApJ, 824, 132

\bibitem[Readhead \& Wilkinson(1978)]{SELFCAL} Readhead, A.~C.~S., \& Wilkinson, P.~N. 1978, ApJ, 223, 25

\bibitem[Smirnov(2011a)]{smirnov2011a} Smirnov, O.~M. 2011a, A\&A, 527, A106

\bibitem[Smirnov(2011b)]{smirnov2011b} Smirnov, O.~M. 2011b, A\&A, 527, A107

\bibitem[Trippe et al.(2010)]{Trippe} Trippe, S., Neri, R., Krips, M., et al.\ 2010, \aap, 515, A40

\end{thebibliography}
\end{document}